On the extended states in the Quantum Hall regime and in zero magnetic field

Hans Nielsen

Ørsted Laboratory, Universitetsparken 5, DK-2100 Copenhagen Ø, Denmark

Fax: (+45) 35 32 04 60, e-mail: hnielsen@fys.ku.dk

Abstract

For a two dimensional electron system in a strong perpendicular magnetic field $B$, there are extended states at each Landau level. We show that if the inverse compressibility is negative, then the extended states float downward in energy when $B$ decreases. We set up a condition for the case where all extended states are below the fermi energy in the low $B$ limit. The condition may explain why a conducting state has been observed in a high mobility Si-Mosfet but so far not in high mobility n-GaAs.

Introduction

Until recently it was believed that all states in a two dimensional electron system are localized [1]. The paradox that extended states exist in a strong perpendicular magnetic field $B$ but not for vanishing $B$ was explained by D.E. Khmelnitskii and R.B. Laughlin [2], who showed that when $B$ decreases the extended states float upward in energy. At $B = 0$ all extended states with energy $E_N$ are then above the fermi energy $E_F$ and all states below $E_F$ are localized. The discovery that an insulator-conductor transition may occur in some 2DES [3] shows that in these systems all extended states cannot go above $E_F$ when $B$ decreases.

The aim of this paper is to extend the above mentioned floating argument [2] to a system of interacting electrons. For noninteracting electrons the chemical potential $E_F$ always increases with the electron density $n$. In agreement with [2] this leads to $\frac{dE_N}{dB} < 0$ which formally expresses that the energy of the $N^{th}$ extended state floats upward for decreasing $B$. If however $E_F$ decreases with increasing $n$, as may be the case in a system of interacting electrons where the negative exchange energy dominates [4], then we get $\frac{dE_N}{dB} > 0$. This means that the extended states float downward in energy with decreasing $B$ and may end below $E_F$. A system with extended states just below the Fermi energy may be a conductor.

The floating argument for interacting electrons.

Following [2] we assume that in a sample, where the size $L$ is of the order of the mean free path $l$, the conductivities are given by Drude-like expressions

$$\sigma_{xx} = ne\mu_t \frac{1}{1+(\mu B)^2} \tag{1a}$$

$$\sigma_{xy} = ne\mu_t \frac{\mu B}{1+(\mu B)^2} \tag{1b}$$

$\mu_t$ is the transport mobility which may be larger than the mobility $\mu = \frac{e\tau}{m^*}$ if the momentum relaxation time is longer than the energy relaxation time. At the scaling $\sigma_{xx} \to 0$ unless $\sigma_{xy} = (N+\frac{1}{2})\frac{e^2}{h}$ which occurs when the Fermi energy $E_F$ coincides with the energy $E_N$ of an extended state at the center of the $N^{th}$ Landau level. The condition that the scaling leads to a finite conductance is then

$$(N+\frac{1}{2})\frac{e^2}{h} = ne\mu_t \frac{\mu B}{1+(\mu B)^2} \tag{2}$$

$$E_N = E_F(n) \tag{3}$$

From (2) (3) we can find:

$$\frac{dE_N}{dB} = \frac{dE_F}{dn} \cdot a_N \cdot \left(-\frac{1}{\mu B^2} + \mu\right) \tag{4}$$

where

$$a_N = \left(N + \frac{1}{2}\right)\frac{e}{h}\mu_t^{-1} \qquad (5)$$

It follows from (2) that $N$ has its maximum value $N = N_{max}$ for $B = \frac{1}{\mu}$. If B decreases from this value then it follows from (4) that

$$\frac{dE_N}{dB} \propto -\frac{dE_F}{dn} \cdot a_N \qquad (6)$$

For the inverse thermodynamic density of state $D_T^{-1} = \frac{dE_F}{dn}$ it is known both from theoretical and experimental investigations [4] that it may be negative for very clean samples (high mobility samples) in an interval $n_0 < n < n_1$ at low densities. Here $n_0$ is the density where $D_T^{-1}$ changes sign from positive to negative. At $n_1$ it becomes positive again. (See Fig. 4 in the paper by J.P. Eisenstein et al. [4]). In such samples all extended states may be below $E_F$ when $B$ decreases to just below $B_{min}$ which is the lowest value (corresponding to $N = 0$) where (2) can be satisfied.

The condition that all extended states are below $E_F$ when $E_0$ (energy of extended states in the first Landau level) passes $E_F$ is

$$E_{N_{max}}(B_{min}) < E_0(B_{min}) \qquad (7)$$

From (2) (3) it follows that (7) can be expressed as

$$E_F\left(n\left(n\mu_t\frac{h}{e}\right)\right) < E_F(n) \qquad (8)$$

(8) is obtained as follows: Generally $E_N(B)$ is found by expressing n from (2) and using it in the left side of (3). For µB << 1 it gives $E_N(B) = E_F\left(\frac{a_N}{\mu B}\right)$ From the definitions $N_{max} + \frac{1}{2} = \frac{h}{e}n\mu_t\frac{1}{2}$ and $\mu B_{min} = \frac{1}{2}\frac{e}{h}\frac{1}{n\mu_t}$ is then follows that (8) is the same as (7). For typical values $n = 10^{15}m^{-2}, \mu_t = 5\frac{m^2}{V \cdot s}$ one has $N_{max} + \frac{1}{2} \approx 10$. A good approximation for $E_F(n)$ is [4]

$$E_F = \frac{1}{D_0}\left(n - \left(\frac{2}{\pi}\right)^{3/2} \frac{1}{r_B} \sqrt{n}\right) \tag{9}$$

where $D_0 = \frac{m^*}{\pi \hbar^2}$ = density of states for free electrons. $r_B = \frac{4\pi\varepsilon\hbar^2}{m^* e^2}$ = Bohr-radius. (9) may be used for $n > n_0$ where $n_0$ is the lowest value where $\frac{dE_F}{dn} = 0$. It is determined by the impurities. From (8) and (9) the condition for extended states below $E_F$ can be calculated to (see appendix)

$$n_0 < n < \frac{0.51}{r_B \sqrt{\mu_t \frac{h}{e}}} \tag{10}$$

According to (9) the density $n_1$ where $\frac{dE_F}{dn}$ shifts from negative to positive is $n_1 = \frac{1}{4}\left(\frac{2}{\pi}\right)^3 \frac{1}{r_B^2}$. For $n_1 > n_2$ where $n_2$ is the upper limit of (10) all extended states end below $E_F$ when $\frac{dE_F}{dn} < 0$. But for $n_1 < n_2$ it may be that only $E_0 < E_F$. However, the other extended states will be very close to $E_F$. From (2) with $N = 0$ we can estimate that at least one extended state exists at $B = \frac{1}{\mu}$ if

$$n > \frac{e}{h}\frac{1}{\mu_t} \equiv n_e \tag{11}$$

In the pioneer experiment [3] with $\mu \approx 5 \frac{m^2}{V \cdot S}$ the transition started at $0.9 \cdot 10^{15} m^{-2}$ and the conductivity σ was increasing with $n$. From (11) $n_e = 0.5 \cdot 10^{15} m^{-2} < n_0$ so the conduction begins at $n_0$, where the extended states start to float down. The upper limit in (10) gives $2.5 \cdot 10^{15} m^{-2}$ for Si in reasonable agreement with the density region investigated in [3]. For n-GaAs where $r_B$ is 5 times larger than for Si and where $\mu_t$ is normally at least 10 times larger the condition (10) means that in n-GaAs the transition may occur at densities there is at least an order of magnitude lower than for Si. Also $n_1 \propto \frac{1}{r_B^2}$ is an order of magnitude lower for n-GaAs than for Si. In a recent experiment on hole-GaAs [5] it was found $\sigma \propto (n-n_e)$ for the conductivity. This Drude-like behaviour seems to give some support to the ideas presented here since the number of extended states is $N \propto (n-n_e)$. When the first extended state passes $E_F$ we expect $\sigma \approx \frac{e^2}{h}$ and $k_F l \approx 1$ as at the minimum conductivity. For $n \geq n_1$ the screening of the impurities may be very effective and eventually supress the weak localization effect. This is the region where Finkel'shtein's [6] ideas may be applied. Finally we mention that negative compressibility means a negative dielectric function. In this case superconductivity may be a possibility [7].

Appendix 1:

Special density values: $n_0$ = density where $\frac{dE_F}{dn}$ shifts from positive to negative. $n_1$ = density where $\frac{dE_F}{dn}$ shifts from negative to positive. $n_2$ = upper limit for region where $E_N < E_0$. $n_e$ = lowest density where one extended state exist. $n_e < n_0$ means that this state is above $E_F$.

Appendix 2:

Derivation of (10): If $a = \mu_t \frac{h}{e}, b = \left(\frac{2}{\pi}\right)^{3/2} \frac{1}{r_B}$ then (8) - (9) gives $an^2 - b\sqrt{a} \cdot n < n - b\sqrt{n}$ from which $an^2 - n = \left(\sqrt{an} + \sqrt{n}\right)\left(\sqrt{an} - \sqrt{n}\right) < b\left(\sqrt{an} - \sqrt{n}\right)$ Then $\sqrt{an} + \sqrt{n} - b < 0$ which is satisfied for $\sqrt{n} \leq \frac{-1 + \sqrt{1 + 4\sqrt{ab}}}{2\sqrt{a}} \approx \frac{\sqrt{\sqrt{ab}}}{\sqrt{a}}$ therefore $n \leq \frac{b}{\sqrt{a}} = \frac{\left(\frac{2}{\pi}\right)^{3/2}}{r_B \sqrt{\mu_t \frac{h}{e}}}$ which is (10).